\documentclass[10pt]{article}
\usepackage{graphicx}

\def\Title#1{\begin{center} {\Large #1 } \end{center}}
\def\Author#1{\begin{center}{ \sc #1} \end{center}}
\def\Address#1{\begin{center}{ \it #1} \end{center}}

\newcommand\pubblock{\rightline{\begin{tabular}{l} Proceedings of the Second Annual LHCP\\ \pubnumber\\
         \pubdate  \end{tabular}}}

\newenvironment{Abstract}{\begin{quotation} \begin{center} 
             \large ABSTRACT \end{center}\bigskip 
      \begin{center}\begin{large}}{\end{large}\end{center} \end{quotation}}

\newenvironment{Presented}{\begin{quotation} \begin{center} 
             PRESENTED AT\end{center}\bigskip 
      \begin{center}\begin{large}}{\end{large}\end{center} \end{quotation}}

\def\beq{\begin{equation}}
\def\eeq#1{\label{#1}\end{equation}}
\def\eeqn{\end{equation}}

\def\beqa{\begin{eqnarray}}
\def\eeqa#1{\label{#1}\end{eqnarray}}
\def\eeqan{\end{eqnarray}}

\let\bar=\overbar

\def\Dslash{\not{\hbox{\kern-4pt $D$}}}
\def\dslash{\not{\hbox{\kern-2pt $\del$}}}

\def\msb{{\bar{\ssstyle M \kern -1pt S}}}

\usepackage{color}
\usepackage{hyperref}
\usepackage{subfig}
\usepackage{graphicx}
\usepackage{caption}
\newcommand\pubnumber{ ATL-PHYS-PROC-2014-111 }

\newcommand\pubdate{\today}

\def\affiliation{
On behalf of the ATLAS Experiment, \\
 II. Physikalisches Institut,
Georg-August-Universit\"{a}t G\"{o}ttingen, \\
Friedrich-Hund-Platz 1, 37077 G\"{o}ttingen, Germany }

\newcommand{\tauhad}{$\tau_{\rm had}$~}

\begin{document}
% \linenumbers

% large size for the first page
\large
\begin{titlepage}
\pubblock

\vfill
\Title{Reconstruction and identification of hadronic decays of tau leptons in ATLAS  }
\vfill

\Author{ Zinonas Zinonos  }
\Address{\affiliation}
\vfill
\begin{Abstract}
Hadronically decaying tau leptons are of prime importance in numerous physics analyses in ATLAS. 
The spectrum of the possible applications of hadronically decaying tau leptons reaches from Standard Model measurements, 
including Higgs searches, to searches for physics beyond the Standard Model. 

The basic principles behind the sophisticated tau reconstruction and identification techniques,
which are specifically designed to identify hadronically decaying taus and reject various background processes, are delineated here
along with current data-driven estimates of their respective performance.
\end{Abstract}
\vfill

% DO NOT CHANGE 
\begin{Presented}
The Second Annual Conference\\
 on Large Hadron Collider Physics \\
Columbia University, New York, U.S.A \\ 
June 2-7, 2014
\end{Presented}
\vfill
\end{titlepage}
\def\thefootnote{\fnsymbol{footnote}}
\setcounter{footnote}{0}
%

% normal size for the rest
\normalsize

\section{Introduction}
Hadronically decaying tau leptons (\tauhad) play an important role in physics analyses in ATLAS~\cite{Aad:2008zzm}. 
Their field of application reaches from Standard Model measurements, including Higgs searches~\cite{ATLAS-CONF-2013-108}, 
to searches for physics beyond the Standard Model~\cite{ATLAS-CONF-2013-090}. 
Therefore, the reconstruction and identification algorithms of hadronically-decaying tau leptons 
have to be continuously adjusted and improved to account for the most up-to-date running conditions 
and to succeed at overcoming the unprecedented experimental challenges faced by the LHC running at higher than ever energies and luminosities.

For the 2012 data-taking period, the \tauhad reconstruction and identification schemes have
been specifically re-optimized for the high number of simultaneous collisions per proton-proton bunch crossing (pile-up).

\section{Reconstruction of Hadronic Tau Decays}

Hadronic tau candidates are reconstructed starting from clusters in the electromagnetic and hadronic calorimeters~\cite{ATLAS-CONF-2013-064}. 
The \tauhad reconstruction is seeded by the anti-$k_t$ jet finding algorithm with a distance parameter of $R = 0.4$. 
A barycenter is formed consisting of the sum of the four-vectors of the constituent topological clusters, assuming zero mass for each of the constituents.
Then, the \tauhad detector axis is calculated by using clusters within $\Delta R = \sqrt{ (\Delta \eta)^2 + (\Delta\phi)^2} < 0.2$ around the barycenter.
The four-vectors of those clusters are recalculated using the tau vertex coordinate system and the vectors are summed up.
Because hadronic tau decays consist of a specific mixture of charged and neutral pions, the energy scale of \tauhad candidates is adjusted independently of 
the jet energy scale. 
The reconstructed energy of \tauhad candidates is calculated to the final energy scale by a simulation-based calibration procedure using clusters, 
within $\Delta R < 0.2$ of the seed jet-object barycentre axis.
Tracks in a cone of radius $\Delta R < 0.2$ from the cluster barycentre, known as the {\it core cone}, are associated to the \tauhad candidate, 
and the electric charge is determined from the sum of the charges of the tracks. Tracks within the {\it isolation annulus}, defined by $0.2 < \Delta R <0.4$ 
of the \tauhad axis, are also counted for variable calculations.

\section{Identification of Hadronic Tau Decays}

\subsection{Discrimination against Jets}

The rejection of jets is provided in a separate identification step using discriminating variables based on tracks with $p_T > 1~\rm GeV$ 
and calorimeter cells found in the core region and in the isolation annulus around the \tauhad candidate direction. 

On average, jets are wider than hadronic tau decays with a given momentum. Therefore, different variables describing the shower shape in both the 
calorimeters and the tracking detectors are used to separate \tauhad from jets, 
with the most important calorimeter shape variable being the fraction of the total tau energy contained in the centermost cone defined by $\Delta R <0.1$.
Important tracking variables are the average $p_T$-weighted track distance from the tau axis, and, in multi-prong decays, 
the distance to the track furthest from the tau axis.
Such discriminating variables are combined in a boosted decision tree (BDT), which is trained separately for 1-prong and 3-prong \tauhad candidates.
Three working points -- labeled as \emph{loose}, \emph{medium} and \emph{tight} -- are defined, corresponding
to target identification efficiencies values of 70\%, 60\% and 40\% for 1-prong and 65\%, 55\% and 35\% for multi-prong \tauhad candidates, respectively.

\begin{figure}%
\centering
\subfloat%[label 1]
 {{\includegraphics[width=0.475\textwidth]{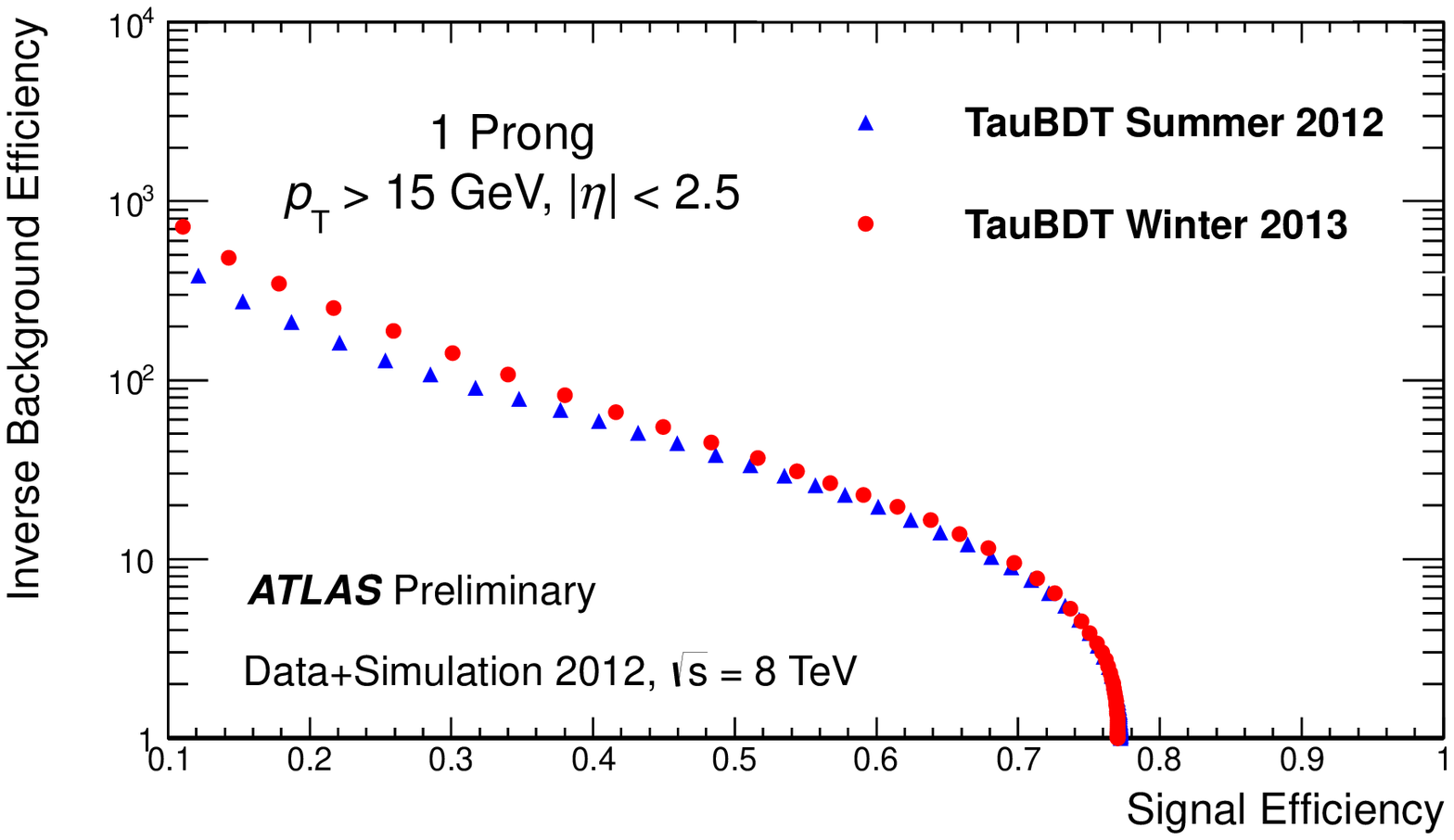} }}%
\hfill
\subfloat%[label 2]
 {{\includegraphics[width=0.475\textwidth]{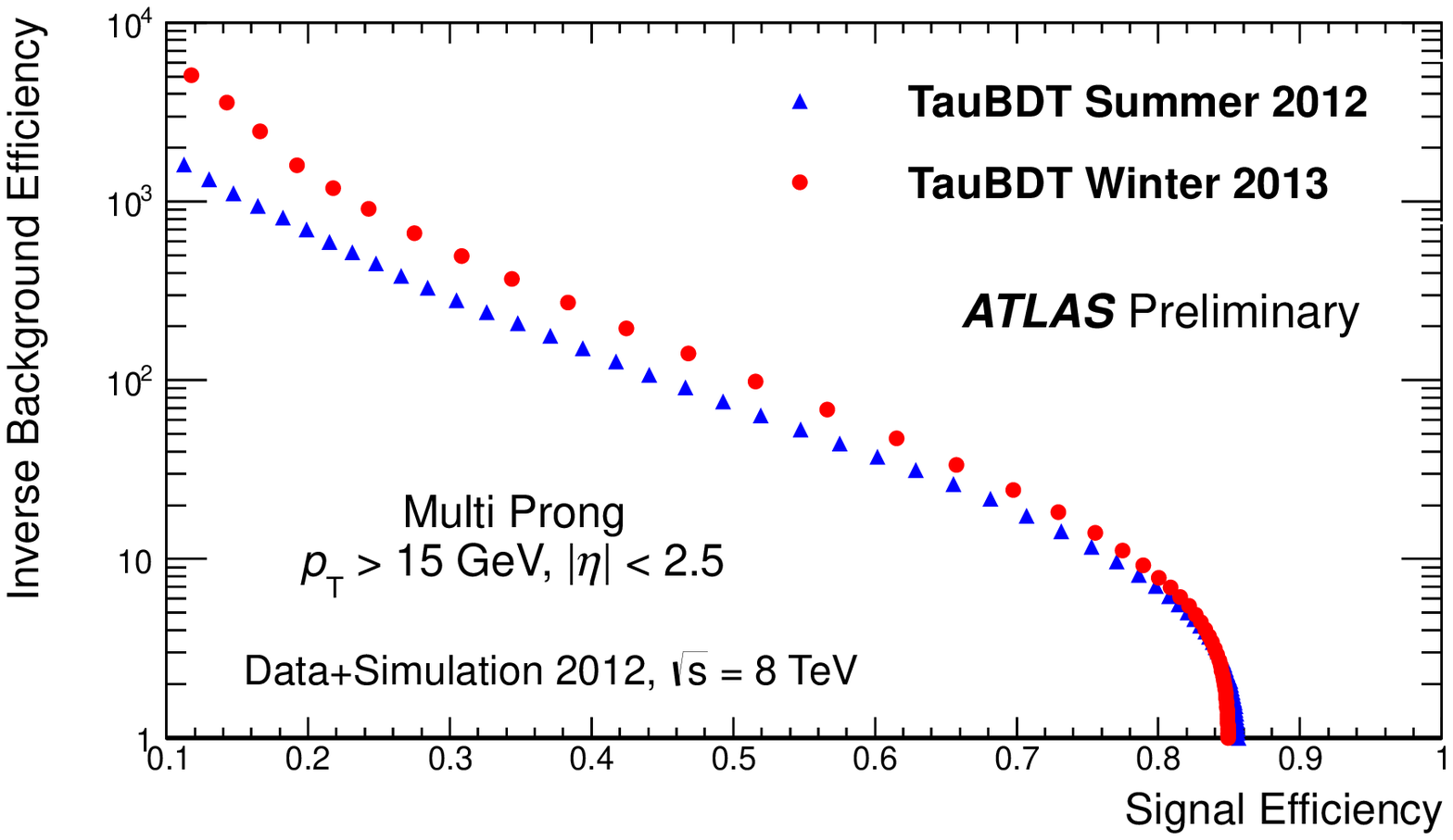} }}%
\caption{Inverse background efficiency as a function of the signal efficiency with a BDT algorithm for 1-prong $\tau_\mathrm{had-vis}$ candidates 
with a $p_\mathrm{T} > 15$ GeV and $|\eta| < 2.5$. 
The signal efficiencies are obtained using $Z\to\tau \tau$, $Z'\to\tau \tau$ and $W\to\tau \nu$ simulated events. 
The background efficiencies are derived using 2012 collision data after applying a multi-jet selection and are calculated with respect to all 
candidates with exactly one reconstructed track. 
The Winter 2013 BDT uses $\pi^{0}$-related variables that increase its performance~\cite{atlas-tau-pub}. }
\label{fig:JetSigBDT}%
\end{figure}

Figure~\ref{fig:JetSigBDT} shows the inverse background efficiency as a function of signal efficiency for the jet BDT discrimination algorithms.
Background rejection factors of $10-40$ for signal efficiencies of 70\% are achieved, going up to 500 for 35\% signal efficiency~\cite{atlas-tau-pub}.

\subsection{Discrimination against Electrons}
\begin{figure}%
\centering
\subfloat%[label 1]
 {{\includegraphics[width=0.45\textwidth]{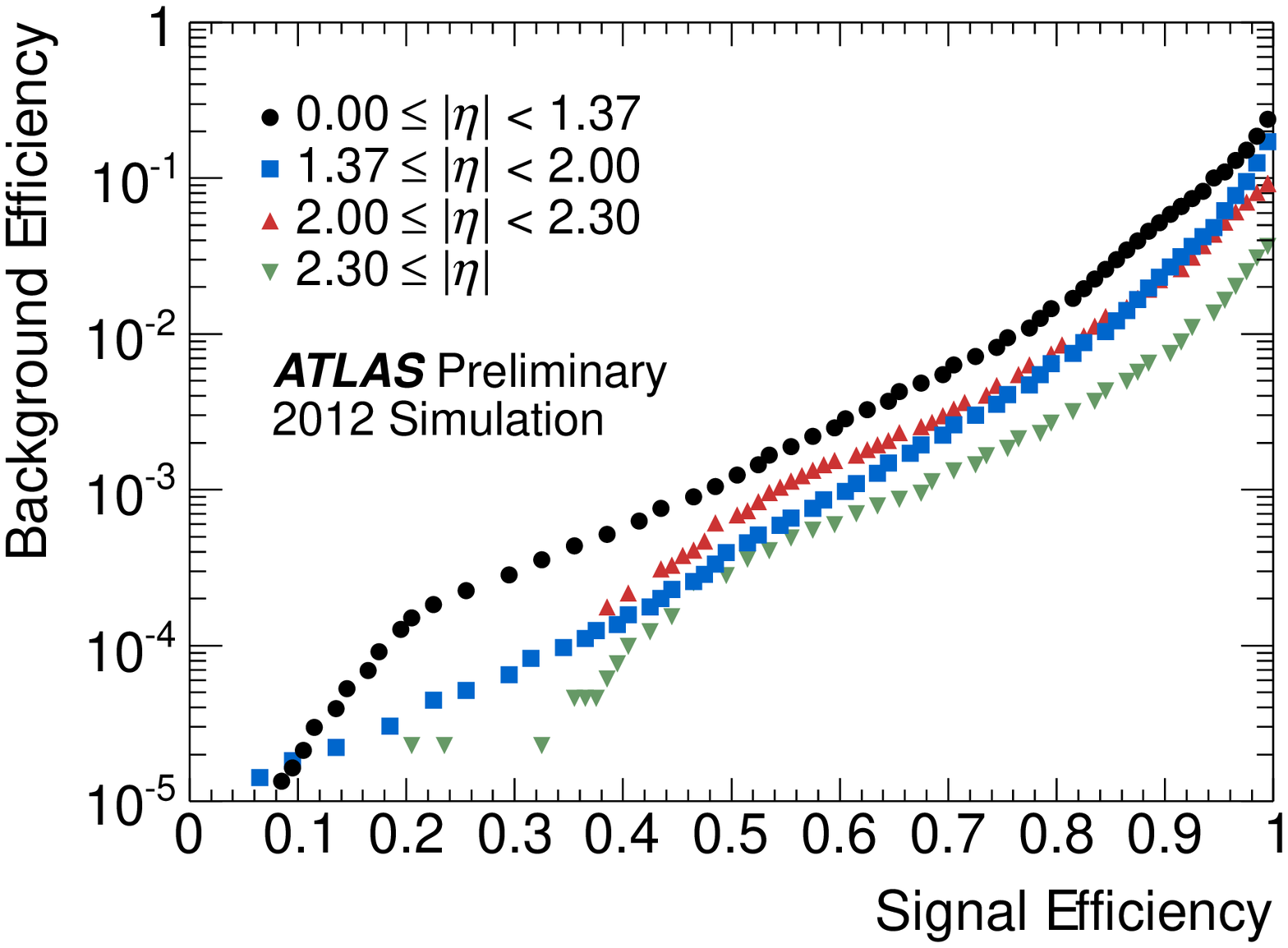} }}%
\hfill
\subfloat%[label 2]
 {{\includegraphics[width=0.45\textwidth]{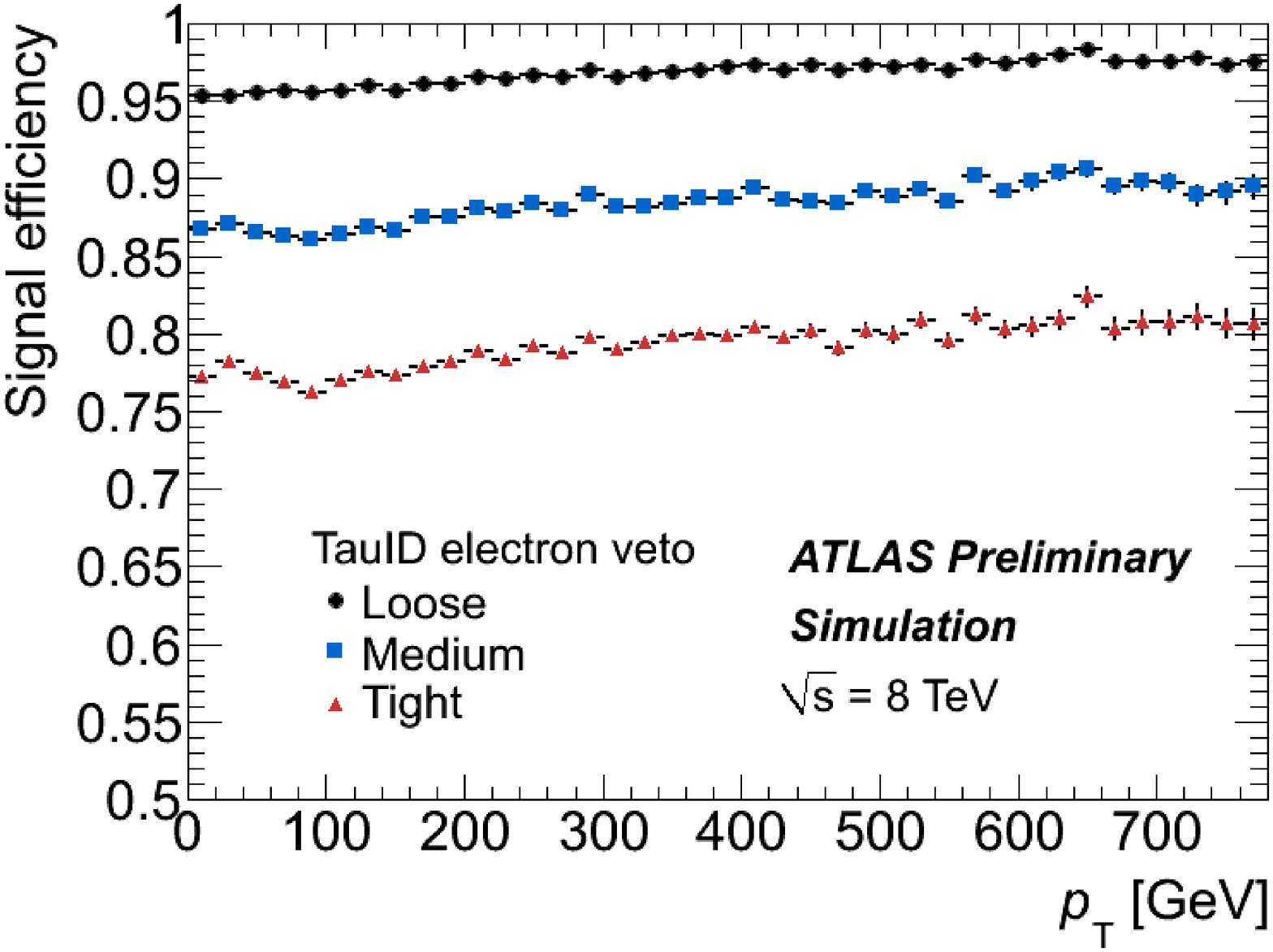} }}%
\caption{Background efficiency as a function of signal efficiencies of the electron veto for each of the different $\eta$ regions (left).
 Signal efficiency of the Winter 2013 electron veto for truth-matched 1-prong tau candidates as a function of reconstructed transverse momentum (right). 
 The efficiencies are obtained using simulated $Z\rightarrow\tau\tau$ events for signal and simulated $Z\rightarrow ee$ events for background~\cite{atlas-tau-pub}.}
\label{fig:EleSigBDT}%
\end{figure}

The characteristic signature of 1-prong \tauhad can be mimicked by electrons.
Despite the similarities of the \tauhad and electron signatures, there are several properties that can be used to discriminate between them. 
The most useful examples of such properties are the emission of transition radiation of the electron track and the longer and wider showers produced by the
hadronic tau decay products in the calorimeter, compared to the one created by an electron.
Such properties are used to define \tauhad identification discriminants specialized in the rejection of
electrons mis-identified as hadronically decaying tau leptons. 
In 2012, the only method for this uses a BDT algorithm.
The electron-veto BDT is optimized using simulated $Z \rightarrow \tau \tau$ events for the signal and 
$Z \rightarrow e e$ events for the background, and was trained for different pseudorapidity ($\eta$) regions.
The signal versus background efficiencies for the different $\eta$ regions and the dependence of the efficiency of the electron veto on $p_T$ are shown 
in Figure~\ref{fig:EleSigBDT}.
The dependence on $p_T$ was such that a sliding cut on the e-veto BDT score was used when implementing the three efficiency working points 
\emph{loose}, \emph{medium} and \emph{tight}. 
These working points were chosen to yield signal efficiencies of 95\%, 85\% and 75\%, respectively.

\subsection{Discrimination against Muons}

As minimum ionizing particles, muons are unlikely to deposit enough energy in the calorimeters to be reconstructed as a \tauhad candidate. 
However, when a sufficiently energetic cluster in the calorimeter is associated with a muon, the muon track and the calorimeter cluster together
may be mis-identified as a \tauhad. 
% The most effective means of rejecting fake \tauhad from muons is to use the default muon recon
% struction algorithms: if a τ had-vis candidate overlaps geometrically with a reconstructed muon it is not
% considered by analyses.
Usually, muons which have deposited significant energy in the calorimeter are likely to have done so mostly in
the hadronic calorimeter, resulting in \tauhad candidates with an unusually low electromagnetic energy fraction.
Also, for such muons the track momentum may be higher than the calorimeter energy, 
which is also true of muons that overlap coincidentally with some other calorimeter deposit.
Finally, the very low-energy muons must overlap with some other calorimeter energy to pass the \tauhad reconstruction, 
and are therefore characterized by a high electromagnetic fraction and low track momentum fraction.

To optimize the muon veto, simulated muons from $Z\rightarrow \mu\mu$ events and hadronic tau decays from $Z \rightarrow \tau\tau$ events were used.
The resulting efficiency is better than 96\% for true \tauhad with a reduction of muon fakes around 40\%~\cite{ATLAS-CONF-2013-064}.

\section{Tau Identification Efficiency Measurement}

The identification (ID) efficiency of \tauhad candidates is studied in data using a 'tag-and-probe' approach.
From these studies, a set of scale factors, defined as $\epsilon_{\mathrm{data} } / \epsilon_{\mathrm{simulation} }$, 
is derived to correct the simulated data samples used in physics analyses.
These measurements are performed independently using $Z \rightarrow \tau \tau$, $W \rightarrow \tau\nu$ and $t \bar{t} \rightarrow \tau + \mathrm{jets}$ events.
Scale factors for the \tauhad mis-identification probability for electrons are also measured, comparing $Z \rightarrow ee$ events in data and simulations.

The tag-and-probe approach chosen consists of selecting events with real tau leptons in their final
state, and extracting a measure of the identification efficiencies directly from the number of reconstructed
\tauhad before and after identification algorithms are applied. 
To estimate the number of background events, a variable with good separation potential is chosen. 
A fit is then performed using the expected distributions in this variable (templates) for both signal and backgrounds. 
%The fit has to be performed multiple times: once before any identification is used and once after each identification algorithm. 
The real number of hadronic tau decays in data is then obtained from the fitted signal template. 
To measure the tau ID efficiency for data, $\epsilon_{\mathrm{data} }$, the number of real \tauhad after tau ID is divided by the number 
of real \tauhad before tau ID. 
The uncertainties on the efficiency measurement are estimated by recalculating the efficiency using systematically altered templates. 
For the efficiency in simulated samples, $\epsilon_{\mathrm{simulation} }$, the number of tau leptons before and after tau ID is taken 
directly from truth-matched \tauhad candidates.
Scale factors are therefore calculated to account for the differences between data and simulation due to the modeling of the input variables 
for the identification algorithms.

The $Z \rightarrow \tau_{\mathrm{lep}} \tau_{\mathrm{had}}$ channel is chosen as the main measurement, 
as it offers the highest precision due to the low associated backgrounds. 

To determine the number of signal and background events in the chosen topologies, a fit to data is used.
The signal and the background are modeled with templates, obtained from either simulations or data-driven techniques. 
The variable chosen for the fit is the extended track multiplicity 
(performing a $p_T$-correlated track counting in the $0.2 < \Delta R < 0.6$ annulus around the core cone) 
associated to the \tauhad candidate as it provides good separation between the signal and the backgrounds.

\begin{figure}%
\centering
\subfloat%[label 1]
 {{\includegraphics[width=0.45\textwidth]{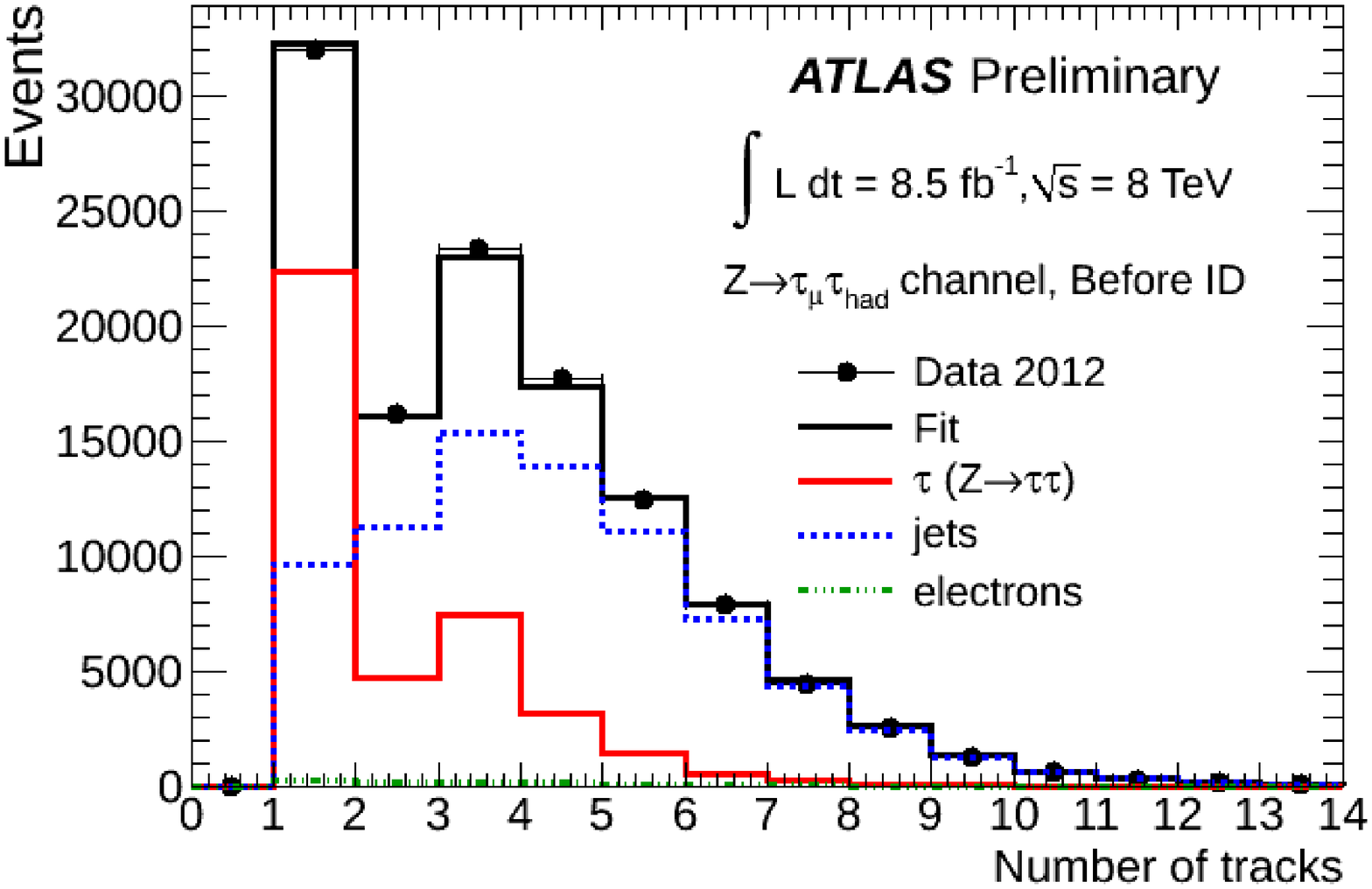} }}%
~
\subfloat%[label 2]
 {{\includegraphics[width=0.45\textwidth]{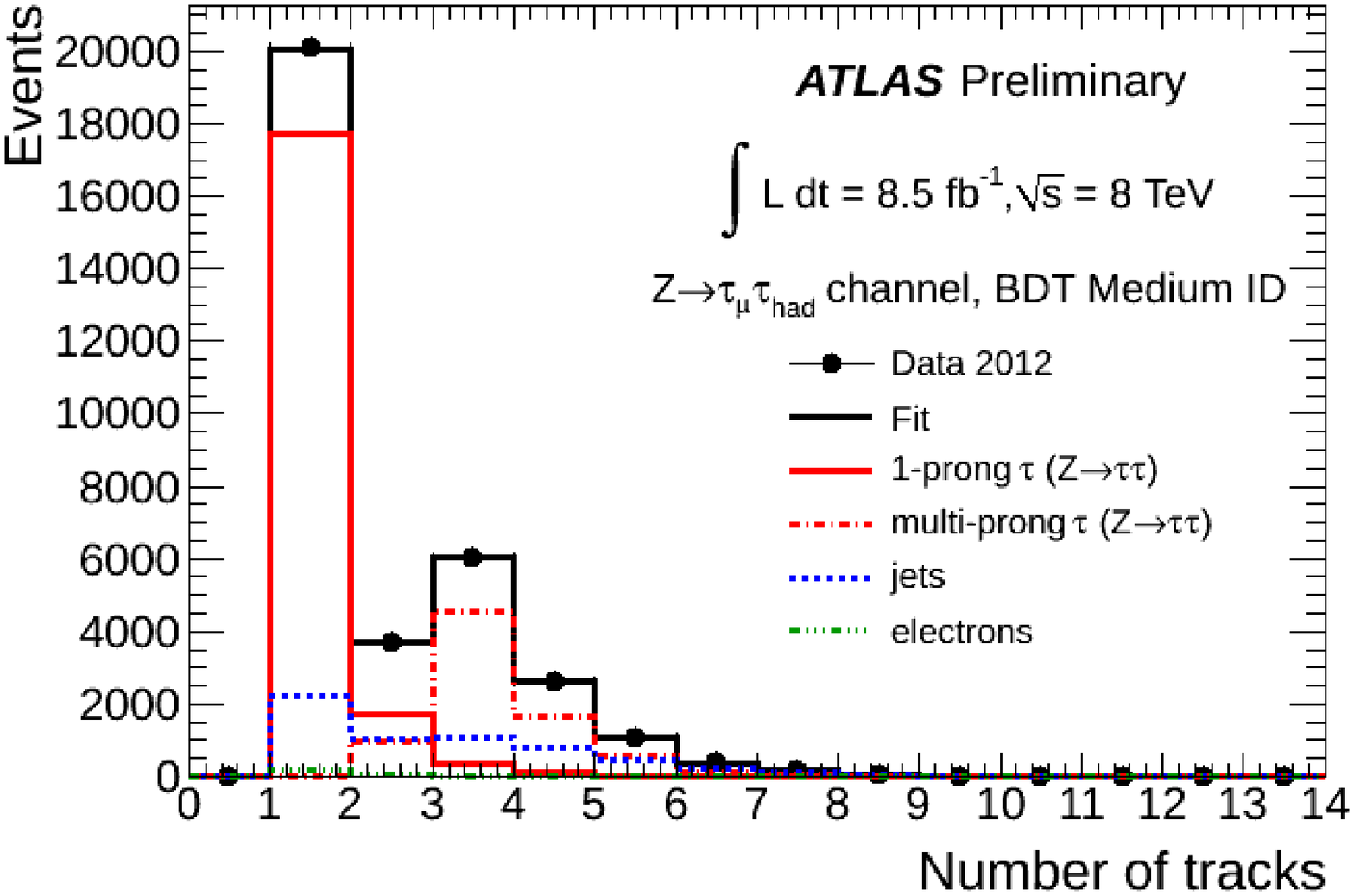} }}%
\caption{Distribution of the number of tracks associated with hadronically-decaying tau candidates from the $Z \to \tau \tau$
tag \& probe selection in 2012 data,  before tau ID (left) and after BDT medium tau ID (right) for
the measurement of the 1-prong and multi-prong identification efficiencies. 
All ID methods and working points are included in the same fit. 
The tau signal template and the electron template are taken from simulations. 
The jet template is obtained from data in a control region. The distribution shown is the sum of core and $p_T$-correlated tracks~\cite{atlas-tau-pub}. }
\label{fig:fit}%
\end{figure}

The extended track multiplicity distribution after the tag-and-probe selection has been applied, and
before and after tau ID criteria have been applied, can be found in Figure~\ref{fig:fit}. 
In these figures, the multi-jet background is not included as it is later determined using data-driven techniques;
this background accounts for most of the difference between the data and simulation.

\section{Conclusions}
The reconstruction and identification algorithms for hadronically decaying tau leptons for the 2012 data period have been outlined here. 
%The variables used in the discriminants have been re-optimized to be robust with the increased amount of pile-up present in the 2012 data. 
%The resulting performance of the identification algorithms shows efficiencies which are independent of the number of simultaneous interactions (pile-up). 
A BDT identification algorithm was trained on simulated event samples and data and provides high background rejection and signal efficiency 
for three pre-determined working points. 
This discrimination algorithm provides rejection factors against multi-jet backgrounds of the order of 60 to 500 for a signal efficiency of 35\%. 
Dedicated prescriptions are also provided for rejecting electrons or muons being mis-identified as \tauhad candidates. 
The electron-veto uses a multi-variate discriminant and was optimized using simulated events, providing three fixed efficiency working points. 
Rejection factors above 100 against background electrons are attained with the tight working point. 
The performance of the various identification algorithms was tested in data-driven studies. 
Measurements of the tau ID algorithm efficiencies were performed using the 
$Z \rightarrow \tau_{\mathrm{lep}} \tau_{\mathrm{had}}$, 
$W \rightarrow \tau_{\mathrm{had}} \nu_\tau$, and
$t \bar{t} \rightarrow \tau_\mathrm{ had} + \mathrm{jets}$ channels with the goal of providing data for simulation scale factors. 

% The resulting inclusive measurements of the scale factors have relative uncertainties at the 2–3\% level.
In general, good agreement was found between the performance of the identification algorithms in both simulated events and in data.
% , with slight divergences appearing at the tighter efficiency working points, where
% the performance is slightly worse in data. Similarly, scale factors were obtained by comparing the ef-
% ficiency of the electron veto algorithm in data and simulated Z → ee events. The performance in the
% central/barrel region of the detector is found to be well-modeled in simulation. In the region |η| > 2.0,
% the calorimeter energy deposits are known to be poorly simulated and a clear discrepancy is observed.
Scale factors and associated uncertainties, to be used in correcting simulated samples, are provided for
the ID algorithms and the electron veto.
%%%%%%%%%%%%%%%%%%%%%%%%%%%%%%%%%%%%%%%%%%%%%%%%%%%%%%%%%%%%%%%%%%%%%%%%%
%%
%%   use this format to include an .eps figure into your paper
%%
% \begin{figure}[htb]
% \centering
% \includegraphics[height=2in]{head_lhcp2014}
% \caption{ Place the caption here}
% \label{fig:figure1}
% \end{figure}
%%%%%%%%%%%%%%%%%%%%%%%%%%%%%%%%%%%%%%%%%%%%%%%%%%%%%%%%%%%%%%%%%%%%%%%%%%%

% See Figure \ref{fig:figure1} and Table \ref{tab:table1}. 

%%%%%%%%%%%%%%%%%%%%%%%%%%%%%%%%%%%%%%%%%%%%%%%%%%%%%%%%%%%%%%%%%%%%%%%%%
%%
%%   use this format to include a LaTeX table  into your paper
%%
% \begin{table}[t]
% \begin{center}
% \begin{tabular}{l|ccc}  
% Patient &  Initial level($\mu$g/cc) &  w. Magnet &  
% w. Magnet and Sound \\ \hline
%  Guglielmo B.  &   0.12     &     0.10      &     0.001  \\
%  Ferrando di N. &  0.15     &     0.11      &  $< 0.0005$ \\ \hline
% \end{tabular}
% \caption{ place the caption here }
% \label{tab:table1}
% \end{center}
% \end{table}
%%%%%%%%%%%%%%%%%%%%%%%%%%%%%%%%%%%%%%%%%%%%%%%%%%%%%%%%%%%%%%%%%%%%%%%%%%%

\end{document}